\documentclass[10pt,twocolumn,pre,floats,english,superscriptaddress,nofootinbib]{revtex4-1}
\usepackage{amsthm}
\usepackage{dsfont}
\usepackage[T1]{fontenc}
\usepackage[latin9]{inputenc}
\usepackage{geometry}
\usepackage{caption,subcaption}
\usepackage{amsmath}
\usepackage{amssymb}
\usepackage{graphicx}
\usepackage{hyperref}
\usepackage{color}
\usepackage{chngpage}

\usepackage{multirow}

\makeatletter
\@ifundefined{textcolor}{}
{%
 \definecolor{BLACK}{gray}{0}
 \definecolor{WHITE}{gray}{1}
 \definecolor{RED}{rgb}{1,0,0}
 \definecolor{GREEN}{rgb}{0,1,0}
 \definecolor{BLUE}{rgb}{0,0,1}
 \definecolor{CYAN}{cmyk}{1,0,0,0}
 \definecolor{MAGENTA}{cmyk}{0,1,0,0}
 \definecolor{YELLOW}{cmyk}{0,0,1,0}
 }
\newcommand{\Tr}{\operatorname{Tr}}
\newcommand{\re}{\operatorname{Re}}

\makeatother

\usepackage{babel}
\begin{document}

\renewcommand\theparagraph{\arabic{paragraph}}

\title{Enumerating Copies in the First Gribov Region on the Lattice in up to four Dimensions}

\author{Dhagash Mehta}

\email{dbmehta@ncsu.edu}

\affiliation{Department of Mathematics, North Carolina State University, Raleigh, NC 27695-8205, USA}
\affiliation{Department of Chemistry, The University of Cambridge, Lensfield Road, Cambridge CB2 1EW, UK.}

\author{Mario Schr\"ock}

\email{mario.schroeck@roma3.infn.it}

\affiliation{Istituto Nazionale di Fisica Nucleare (INFN), Sezione di Roma Tre, Rome, Italy}
\begin{abstract}
The covariant gauges are known to suffer from the Gribov problem: even after fixing a gauge non-perturbatively, there may still exist residual copies which are physically equivalent to each other, called Gribov copies. While the influence of Gribov copies in the relevant quantities such as gluon propagators has been heavily debated in recent studies, the significance of the role they play in the Faddeev--Popov procedure is hardly doubted. We concentrate on Gribov copies in the first Gribov region, i.e., the space of Gribov copies at which the Faddeev--Popov operator is strictly positive (semi)definite. We investigate compact U($1$) as the prototypical model of the more complicated standard model group SU($N_{c}$). With our Graphical Processing Unit (GPU) implementation of the relaxation method we collect up to a few million Gribov copies per orbit. We show that the numbers of Gribov copies even in the first Gribov region increase exponentially in two, three and four dimensions. Furthermore, we provide strong indication that the number of Gribov copies is gauge orbit dependent.
\end{abstract}
\maketitle

\section{Introduction}
The most successful way of studying gauge field theories non-perturbatively is to put them on a finite space-time lattice, this approach is commonly referred to as lattice field theory ~\cite{Rothe:2005nw}. In the continuum, promising non-perturbative approaches are
functional methods, in particular Dyson-Schwinger equations (DSEs)~\cite{Alkofer:2000wg} and functional renormalization group equations (FRGs).
The DSE approach, for example, can be useful in the low momentum region
of quantum chromodynamics (QCD), whereas using lattice QCD one can perform first
principles calculations of non-perturbative quantities in QCD. The
approximations involved in lattice QCD can be systematically removed\footnote{In practice, due to limited computer power, an extrapolation to the infinite volume and continuum limits has to be performed of which the analytical form is unknown.},
whereas a systematic removal of the truncations in DSEs is much more involved.
In other words, lattice simulations can provide
an independent check on the results obtained in the DSE approach.

There is however a subtle difference in the two approaches. A lattice field theory is manifestly gauge invariant, hence one does not need to fix a gauge on the lattice to calculate
gauge invariant observables. In the continuum approaches, each gauge configuration comes with
infinitely many equivalent physical copies, the set of which is called
a gauge-orbit. Hence, to remove the redundant degrees of freedom, one requires gauge fixing. Thus to compare with DSE results,
a corresponding gauge fixing is necessary on the lattice.

In the continuum, the standard way to fix a gauge in the perturbative limit is
the so-called Fad\-de\-ev--Popov (FP) procedure~\cite{Faddeev:1967fc} which amounts to formulate a gauge fixing device which is called the gauge fixing partition function,
$Z_{GF}$. In the perturbative limit, it can be shown that for an ideal gauge fixing condition, $Z_{GF} = 1$. Then, this unity is inserted in the measure of the generating
functional so that the redundant degrees of freedom are removed after
appropriate integration. Becchi, Rouet, Stora and Tyutin (BRST) generalised the FP procedure~\cite{Becchi:1975nq}.

Gribov found that in non-Abelian gauge theories
a generalised Landau gauge fixing condition, if treated non-perturbatively,
has multiple solutions, called Gribov or Gribov--Singer copies~\cite{Gribov:1977wm,Singer:1978dk,Alkofer:2000wg}. Thus, the above assumption of the ideal gauge fixing condition became a subtle point in generalizing the FP procedure for non-perturbative field
theories. Furthermore, Neuberger showed that on the lattice,
the corresponding $Z_{GF}=0$ ~\cite{Neuberger:1986vv,Neuberger:1986xz}, i.e., the expectation value of a gauge fixed observable awkwardly turns out
to be $0/0$, known as the
Neuberger $0/0$ problem. It yields that BRST formulations can not be constructed on the
lattice, a situation which may severely hamper any comparison of gauge-dependent quantities on the lattice with those in the continuum.

It is argued that
Gribov copies may influence the infrared behaviour of the gauge dependent propagators
of gauge theories both on the lattice \cite{Aouane:2011fv,Bornyakov:2011fn,Cucchieri:1997dx},
and in the continuum \cite{Ilderton:2007dh,Ilderton:2007qy}. In \cite{Maas:2009ph} SU(2) Yang--Mills theory has been investigated in the strong coupling limit on the lattice: it has been shown that the Gribov ambiguity is rather strong in that case and especially affects the ghost propagator.
There have also been efforts to count Gribov copies in the continuum in Refs.~\cite{Holdom:2007gx,Holdom:2009ws} where the counting was restricted for the static spherically
symmetric configurations only, for the SU($2$) case. Interestingly, recently, a deep relation between lattice gauge fixing and lattice supersymmetry has been proposed \cite{Catterall:2011aa,Mehta:2011ud}: the partition functions of a class of supersymmetric Yang-Mills theories can be viewed as a gauge fixing partition function a la Faddeev--Popov and the `Gribov copies' are then nothing but the classical configurations of the theory.

\paragraph{Landau gauge on the lattice:}
On the lattice, gauge fixing is reformulated as an optimization problem. With the gauge fields defined through link variables $U_{i,\mu}\in G$
where the discrete variable $i$ denotes the lattice-site index, $\mu=1,\hdots,d$ is a directional
index and $G$ is the corresponding group of the theory. The standard choice of the lattice Landau
gauge (LLG) fixing functional to be optimized with respect to the
corresponding gauge transformations $g_{i}\in G$, is
\begin{equation}
F_{U}(g)=\sum_{i,\mu}(1-\frac{1}{N_{c}}\re \Tr g_{i}^{\dagger}U_{i,\mu}g_{i+\hat{\mu}}),\label{eq:general_l_g_functional}
\end{equation}
for SU($N_{c}$) gauge groups. Choosing $f_{i}(g):=\frac{\partial F_{U}(g)}{\partial g_{i}}=0$
for each lattice site $i$ gives the lattice divergence of the lattice
gauge fields and in the naive continuum limit recovers the continuum Landau gauge condition, i.e., $\partial_{\mu}A_{\mu}=0$, where $A_{\mu}$ is the gauge potential. The corresponding FP operator $M_{FP}$ is then the
Hessian matrix of $F_{U}(g)$ with respect to the gauge transformations. The stationary points of $F_{U}[g]$ are the Gribov copies.

Neuberger showed~\cite{Neuberger:1986vv,Neuberger:1986xz} that when
all the stationary points of $F_{U}[g]$ are taken into account, the gauge fixing partition function $Z_{GF}$  \`{a} la FP procedure turns out to be zero and the expectation
value of a gauge fixed variable is then $0/0$. The Morse theory interpretation of this problem was given by Schaden \cite{Schaden:1998hz} who showed that $Z_{GF}$
calculates the Euler character $\chi$ of the group manifold $G$
at each site of the lattice. In particular, for a lattice with $N$ lattice sites,
\begin{equation}
Z_{GF}=\sum_{i}\mbox{sign}(\det\, M_{FP}(g))=(\chi(G))^{N},\label{eq:Z_GF_is_Euler_char}
\end{equation}
where the sum runs over all the Gribov copies. Since $\chi$ of the group
manifold for compact U(1), $S^1$, and of the group manifold of SU($N_c$), $S^3\times S^5 \times \dots \times S^{2 N_c-1}$, is zero, the corresponding $Z_{GF} = 0$.

To evade this problem, for an SU(2) gauge theory, Schaden proposed
to construct a BRST formulation only for the coset space SU(2)/U(1)
for which $\chi\neq0$. The procedure can be generalised to fix the gauge of
an SU($N_{c}$) lattice gauge theory to the maximal Abelian subgroup
$($U(1)$)^{N_{c}-1}$ \cite{Golterman:2012ig,Golterman:2004qv}. Thus, the Neuberger $0/0$ problem for an SU($N_{c}$)
lattice gauge theory actually lies in (U(1)$)^{N_{c}-1}$. For this reason, we concentrate on the compact
U(1) case in the rest of the paper.

There are other ways proposed to avoid the Neuberger $0/0$ problem by modifying the gauge fixing
condition while taking into account that the corresponding $Z_{GF}$ should be orbit-independent, and, for technical convenience, it should be possible to efficiently implement the corresponding gauge fixing numerically.
Renormalization, in contrast, is not required: unitary gauge in gauge-Higgs models, for example, is even perturbatively non-renormalizable but still yields the correct physics.

In minimal lattice Landau gauge, one focuses on the first Gribov region \cite{Zwanziger:1989mf}, i.e., the space of minima, in which there is no cancelation among the signs of $M_{FP}$. Hence, $Z_{GF}$ just counts the number of minima of $F_{U}[G]$, and the Neuberger
$0/0$ is avoided. It is yet to be shown if the
corresponding $Z_{GF}$
is orbit-independent in general. However, in the one-dimensional \cite{Mehta:2009, Mehta:2010pe} and two-dimensional~\cite{Hughes:2012hg} compact U($1$) cases, it was already shown that $Z_{GF}$ is in fact an orbit-dependent quantity. In the
present paper, one of our goals is to verify this in higher dimensional
cases.

In absolute lattice Landau gauge, one focuses on the space of global minima, called the
fundamental modular region (FMR). The Neuberger $0/0$ problem is again avoided here. It is anticipated that there are
no Gribov copies inside the FMR \cite{Zwanziger:1993dh,vanBaal:1997gu}, which was verified to be true in one- and two-dimensional compact U($1$) cases \cite{Mehta:2009,Mehta:2010pe, Hughes:2012hg}.

Other approaches to evade the Neuberger $0/0$ problem were recently put forward in \cite{vonSmekal:2007ns,vonSmekal:2008es,vonSmekal:2008ws,Testa:1998az,Maas:2009se,Maas:2013vd,Kalloniatis:2005if,Ghiotti:2006pm,vonSmekal:2008en,Mehta:2014orbifold} and reviewed in \cite{Maas:2011se}.

\paragraph{Lattice Landau gauge for compact U(1):}
Following the notations of Ref. \cite{Hughes:2012hg}, for compact U(1) the gauge fields and gauge transformations
are $U_{i,\mu}=e^{i\phi_{i,\mu}}$ and $g_{i}=e^{i\theta_{i}}$, respectively, where the angles $\theta_i$ and $\phi_{i,\mu}$ take values from $(-\pi,\pi]$. Hence, Eq.~(\ref{eq:general_l_g_functional})
becomes
\begin{equation}
F_{\phi}(\theta)=\sum_{i,\mu}\big(1-\cos(\phi_{i,\mu}+\theta_{i+\hat{\mu}}-\theta_{i})\big)\equiv\sum_{i,\mu}(1-\cos\phi_{i,\mu}^{\theta}),\label{eq:sllg-functional}
\end{equation}
where $\phi_{i,\mu}^{\theta}:=\phi_{i,\mu}+\theta_{i+\hat{\mu}}-\theta_{i}$.

When $\phi_{i,\mu}$ is picked randomly, it is refereed to as a random or \emph{hot} orbit and when all the $\phi$-angles are zero, it is called the trivial or \emph{cold} orbit.

We concentrate on periodic boundary conditions (PBC), i.e., $\theta_{i+N\hat{\mu}}=\theta_{i}$ and $\phi_{i+N\hat{\mu},\mu}=\phi_{i,\mu}$, which is the most natural choice in lattice gauge
theories. We remove the global gauge 
degree of freedom by fixing the angle $\theta_{(N,...,N)}$ to zero.
Furthermore, we let $\{\phi_{i,\mu}\}$ take random values independent of
the action, corresponding to the strong coupling limit $\beta=0$, which is sufficient to answer the basic questions of counting Gribov copies and their orbit-dependence as every gauge orbit has a non-vanishing weight for any finite $\beta$. 
The global minimum of $F_{\phi}(\theta)$ is usually thought to be unique modulo
possible accidental degeneracies which are expected to form a set of measure
zero (and non-accidental degeneracies on the boundary of the FMR). Therefore, we focus on the minimal lattice Landau gauge in this work.

\section{What is known So Far}
All the Gribov copies for the one-dimensional LLG for compact U($1$)  have been found analytically for periodic \cite{Mehta:2009,Mehta:2010pe} and antiperiodic \cite{Mehta:2009,vonSmekal:2007ns,vonSmekal:2008es} boundary conditions. However, solving the stationary equations
in more than one dimension turns out to be a difficult task and has
not been done so far. The main difficulty here is that the stationary
equations are highly nonlinear in higher dimensions. In Ref. \cite{Mehta:2009}
it was shown how these equations could be
viewed as a system of polynomial equations, and then the numerical polynomial homotopy continuation method (NPHC) was used to
find all the stationary points for small lattices in two dimensions.
The method was used extensively afterwards to study similar problems of finding stationary points or minima of a multivariate function arising in statistical mechanics and particle
physics \cite{Mehta:2009zv,Hughes:2012hg,Mehta:2011xs,
Mehta:2011wj,Kastner:2011zz,Nerattini:2012pi,
Mehta:2012qr,Maniatis:2012ex,Mehta:2012wk,
Hauenstein:2012xs,Greene:2013ida,Mehta:2013fza,MartinezPedrera:2012rs,He:2013yk}.
Interestingly, in Ref.~\cite{Nerattini:2012pi}, two types of singular solutions were
observed for the trivial orbit case: isolated singular solutions at which the Hessian matrix is singular
(these solutions are in fact multiple solutions) and a continuous family of singular solutions. It was shown that one can
construct one-, two-, etc. parameter solutions, even after fixing
the global $O(2)$ freedom.

The authors of \cite{deForcrand:1994mz} studied the continuum limit of lattice U($1$) theory in two dimensions and found that in that limit, the absolute and local minima become more and more degenerate.

In Ref. \cite{Hughes:2012hg}, for the two-dimensional case, among
other results using the Conjugate Gradient method it was conjectured
that the number of Gribov copies in the first Gribov region increases exponentially. In Ref. \cite{Mehta:2013iea,Mehta:2014appear}, the problem of finding minima of the compact U($1$) LLG in two dimensions was studied for the trivial orbit case, which is nothing but the two-dimensional XY model without disorder. There, many minima were found using potential energy landscape methods \cite{Wales:04,RevModPhys.80.167}, and it was shown that the number of minima increased exponentially in this case. Moreover, using disconnectivity diagrams, it was shown how the minima were connected to each other via the saddles of index 1 (called transitions states in theoretical chemistry). In the current paper, we want to verify this conjectured exponential increase in three and four dimensions. As a byproduct, we
also improve on the previous results for two dimensions. In a separate work, we develop a novel and efficient method to find many Gribov copies, if not all, 
starting from a maximum of the lattice Landau gauge fixing functional and moving towards lower index saddles \cite{Mehta:2014accepted}.

\begin{figure}[htb]
\includegraphics[width=0.5\textwidth]{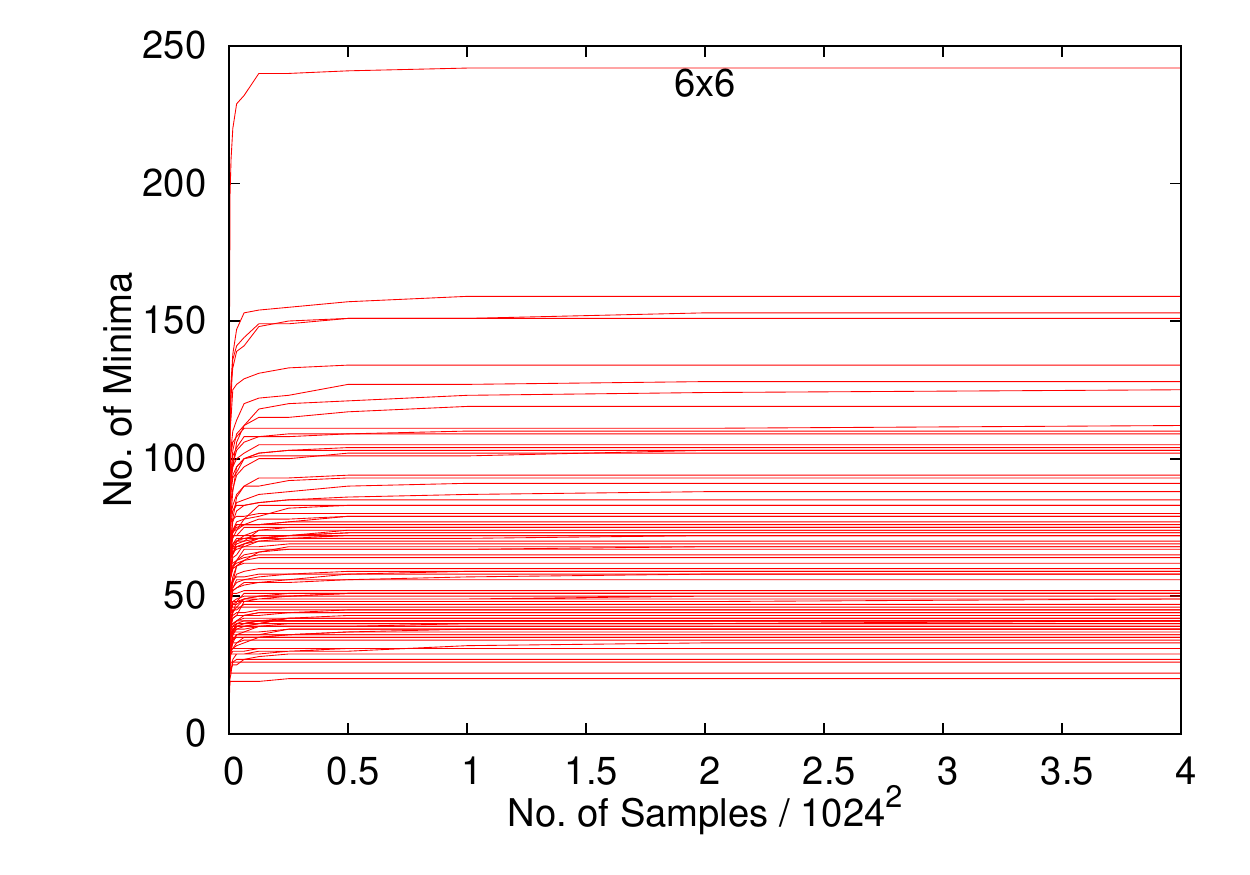}\\
\includegraphics[width=0.5\textwidth]{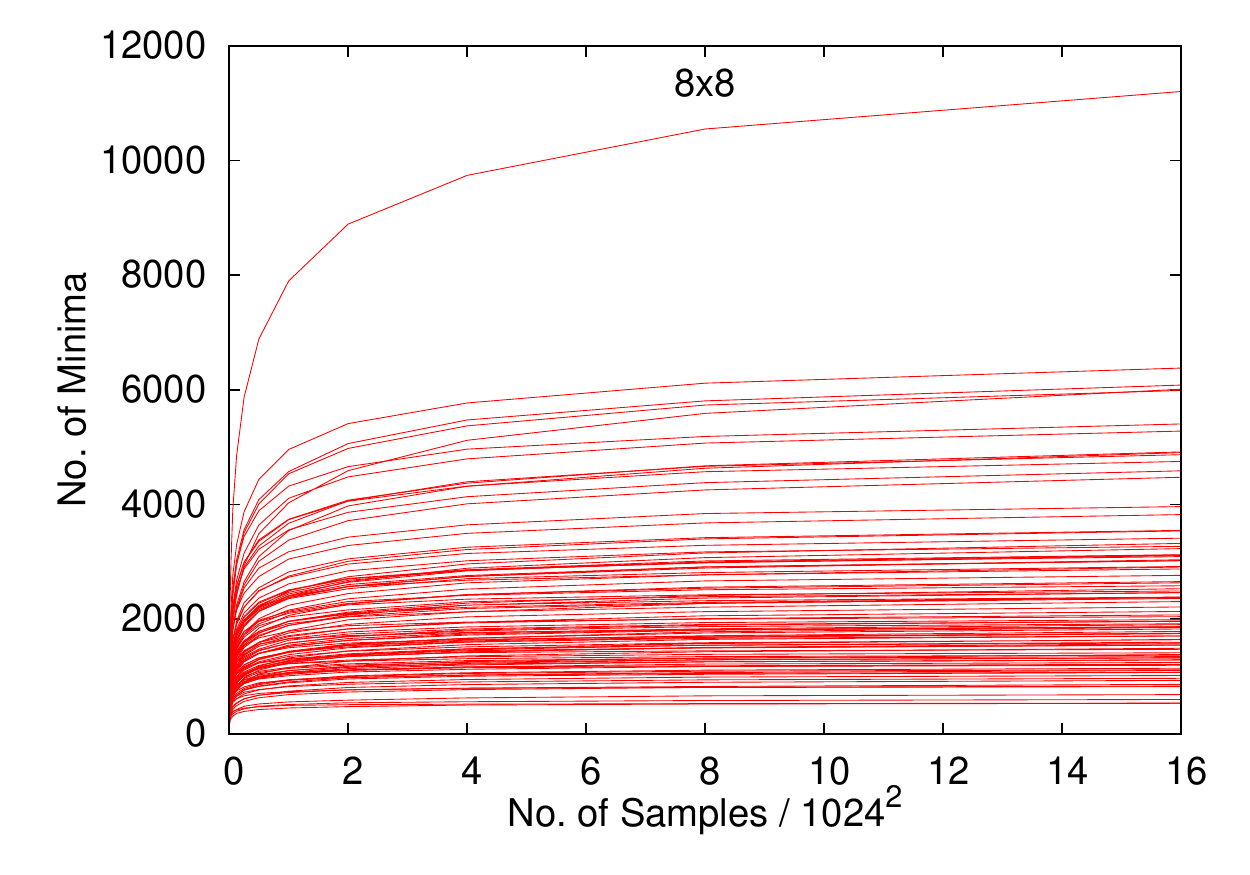}\\
\includegraphics[width=0.5\textwidth]{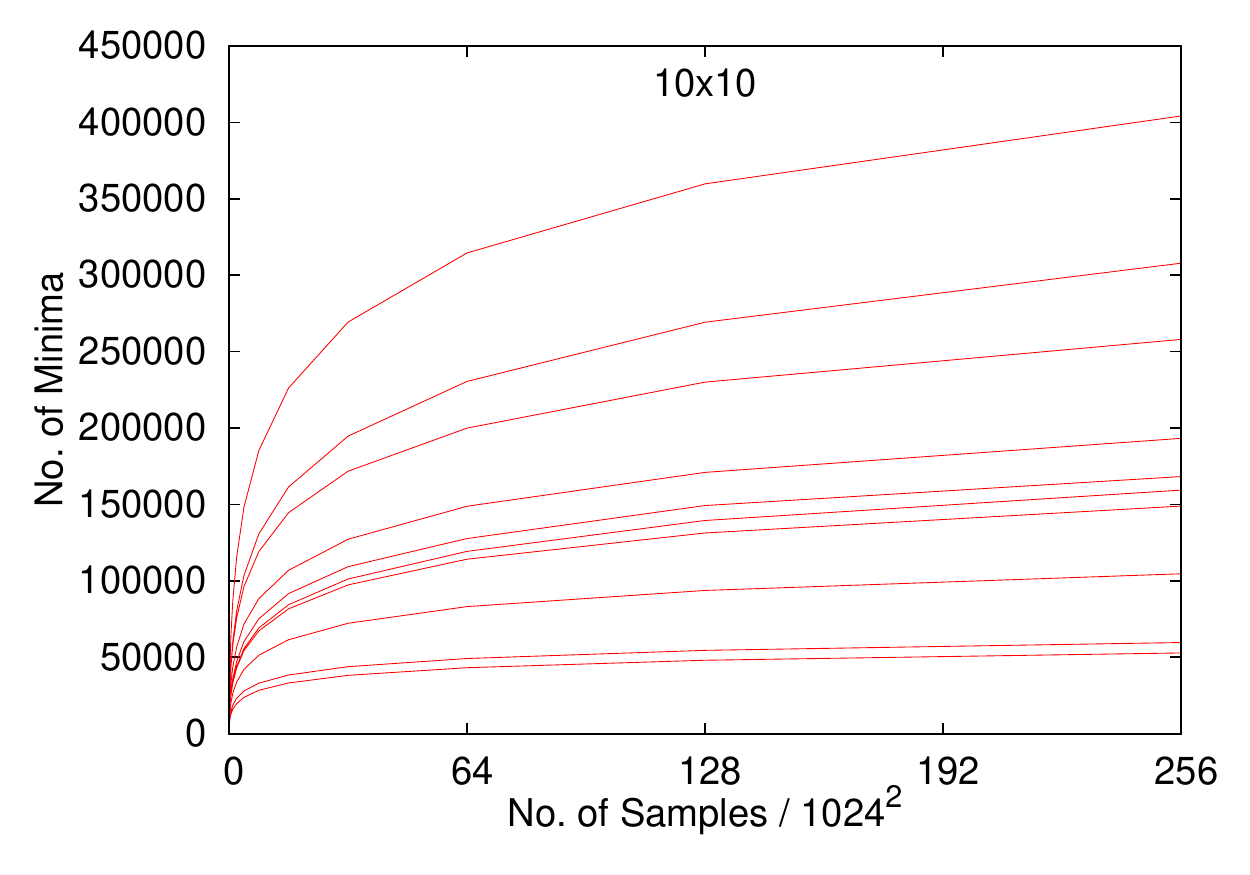}
\caption{The number of distinct minima versus the number of samples (initial guesses) for different orbits for $d=2$ and $N=6, 8, 10$.}
\label{fig:nMinima_vs_nSamples_dim2}
\end{figure}

\begin{figure}[htb]
\includegraphics[width=0.5\textwidth]{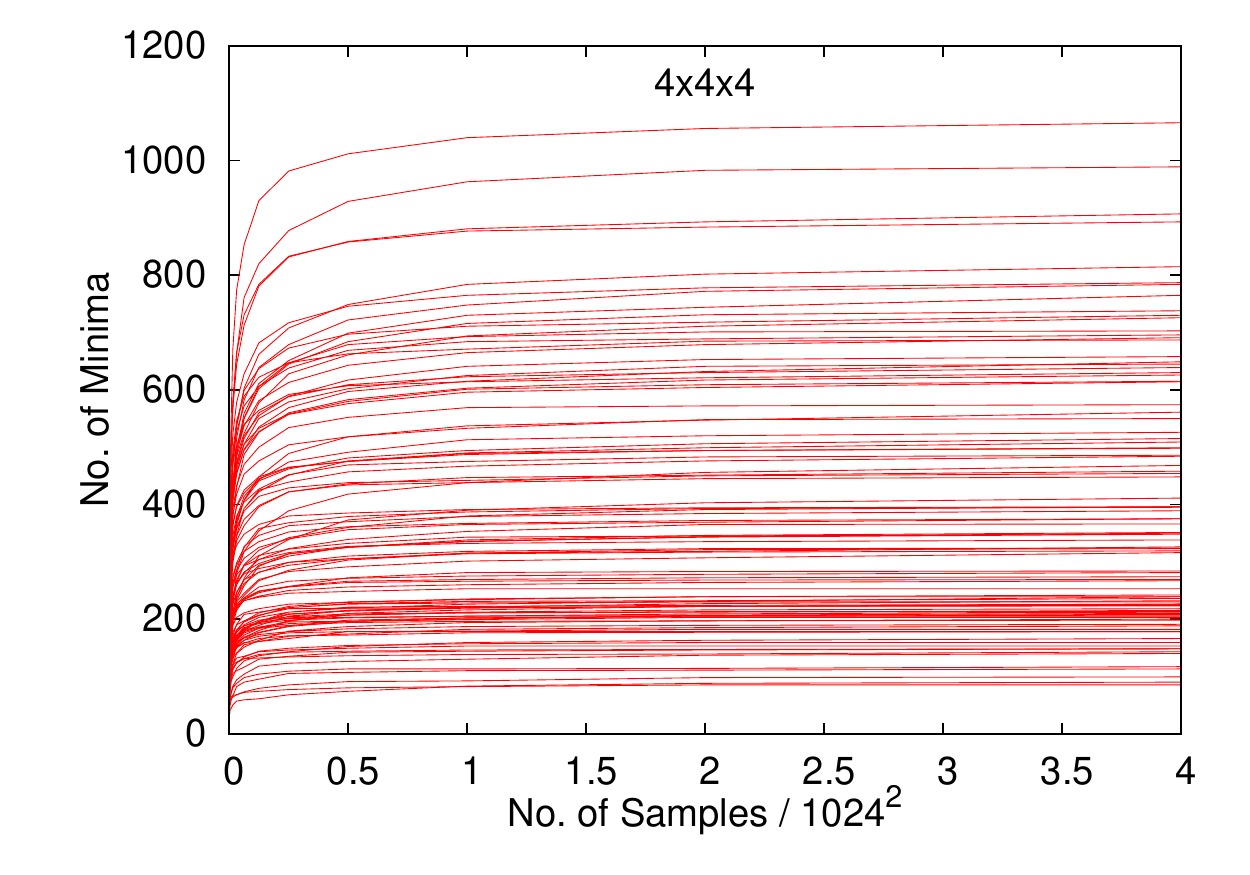}\\
\includegraphics[width=0.5\textwidth]{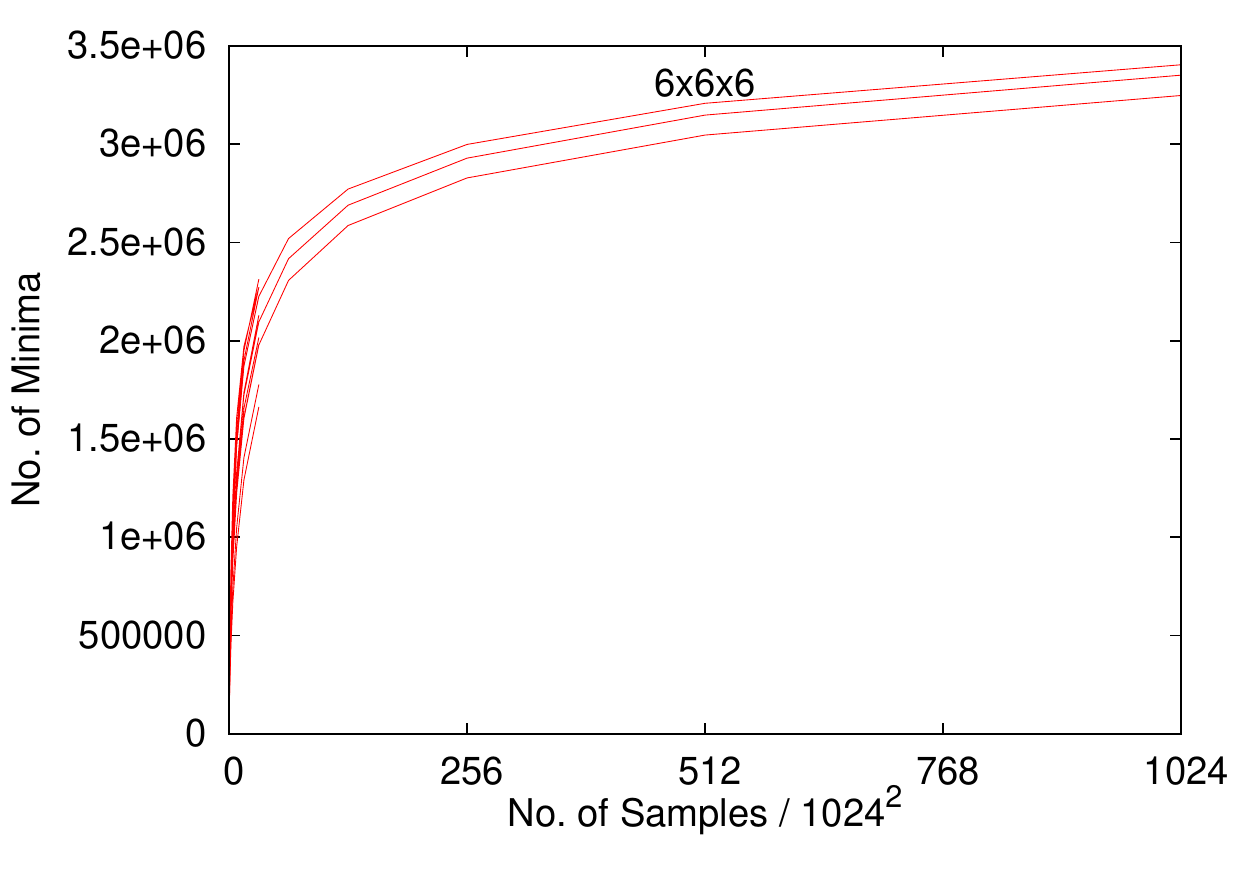}
\caption{The number of distinct minima found versus the number
of samples (random starts) for the three different orbits for $d=3$ and $N=4, 6$.}
\label{fig:nMinima_vs_nSamples_dim3}
\end{figure}

\begin{figure}[htb]
\includegraphics[width=0.5\textwidth]{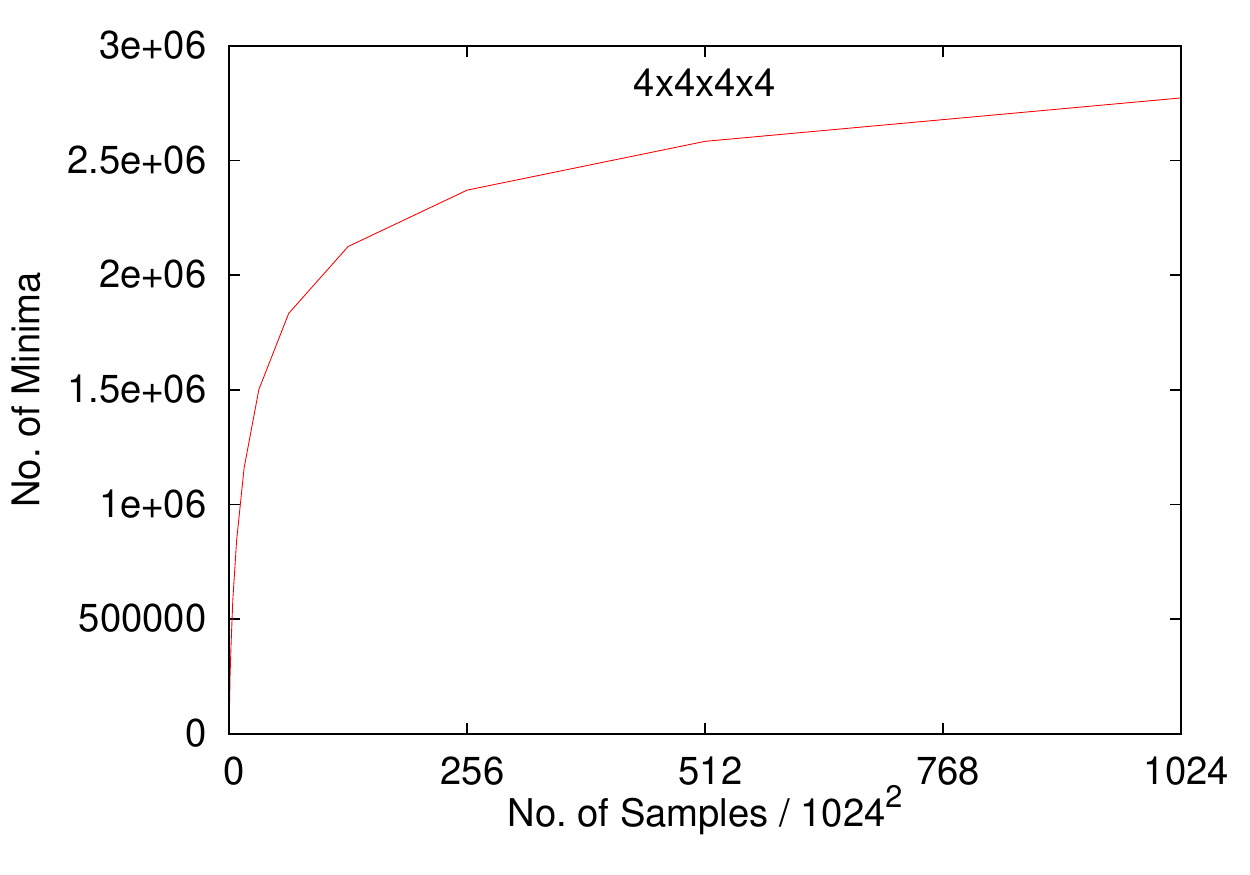}
\caption{The number of distinct minima found versus the number
of samples (random starts) for a single orbit under investigation for $d=4$ and $N=4$}
\label{fig:nMinima_vs_nSamples_dim4}
\end{figure}

\section{A GPU Implementation of the Relaxation Method}
In this section we describe our numerical approach. We adopt the relaxation algorithm
and execute it on graphics processing units (GPUs) which offer a high level
of parallelism and thus enable us to gather a large number of samples within
a practical amount of computer time.

The idea of the relaxation algorithm is to sweep over the lattice while optimizing
the gauge functional (\ref{eq:general_l_g_functional}) locally on each lattice site. Thus, on each site $i$ the maximum of
\begin{equation}\label{eq:maxK}
\frac{1}{2}\mathrm{Re}\left[ g_i K_i \right]
\end{equation}
is to be found. Here we introduced
\begin{equation}
K_i \equiv \sum_\mu
\left[ U_{i,\mu}\;g_{i+\hat\mu}^\dagger + U_{i-\hat\mu,\mu}^\dagger\; g_{i-\hat\mu}^\dagger \right].
\end{equation}
It is easy to see that
Eq.~(\ref{eq:maxK}) becomes maximized if we simply set $\theta_i$ equal to the phase of $K_i^*$.

Note that the local optimization depends for each lattice site on the nearest neighbors only,
hence we can perform a checker board decomposition of the lattice and the local optimization
of all lattice sites of one of the two sublattices will be independent of all
other lattice sites of the same sublattice.
Our implementation will benefit therefrom by optimizing all lattice sites
of a given sublattice concurrently  instead of performing serial loops over
the members of each sublattice.
We do not perform overrelaxation since we have found that, while overrelaxation decreases the convergence time, it also decreases the chance of converging to larger minima and thus introduces a bias.

NVIDIA offers with \emph{CUDA} (Compute Unified Device Architecture) 
a parallel programming model that enables the programmer to run so-called kernels on the GPU.
These kernels are specialised functions that perform a sequence of tasks in a highly
parallel fashion. The user defines a grid of thread blocks and a number of threads per thread block
in such a way that the kernel call replaces serial loops over memory addresses by concurrent
calculations on all corresponding addresses.

Our implementation is based on the \emph{cuLGT}\footnote{\url{http://www.cuLGT.com}} code for lattice gauge fixing on GPUs \cite{Schrock:2012fj}.
Here we assign one thread to one lattice site and a whole lattice will be encapsulated
in one thread block. The GPU can handle several thread blocks per multiprocessor concurrently
and we launch a grid of thread blocks where each thread block contains a lattice
initialised with random numbers which we generate with the \emph{Philox} random number generator \cite{Salmon:2011:PRN:2063384.2063405}. In this way we minimize in parallel as many samples as we launch thread blocks.
Moreover, we add another layer of parallelism by adopting multiple GPUs: therefore we loop over the
kernel calls in the main function of the execution code while switching between the multi-GPUs.

In practice we adopt four cards of the NVIDIA Tesla C2070 and we launch 1024 thread blocks
(i.e. random start samples) per GPU. Hence we run 4096 samples at once.
While CUDA allows for a much larger number of thread blocks to be launched per kernel call
this can be counterproductive since the runtime depends on the slowest converging sample
among all running samples. A smaller number than 1024 thread blocks per GPU, on the other hand,
would not fully occupy the GPU and thus result again in a performance loss.
Therefore, we keep the grid size as 1024 blocks per GPU fixed.
For each sample we store the value of the minimum to which the relaxation algorithm
has converged and subsequently sort these values via bitonic sort, again accelerated by the GPU.

The execution time of the code depends on the lattice size and the number of iterations until convergence which varies from sample to sample. In practice, the time to minimize one \emph{mebisample} ($1024^2$ samples) adopting four NVIDIA Tesla C2070 GPUs varies from a few seconds (e.g. $d=2$ up to $N=6$ lattices) over several minutes (e.g. $10^2$ and $4^3$) up to a few hours ($8^3$ and $4^4$).

As a stopping criterion we require
the largest gradient over all lattice sites and all concurrently running samples to be smaller than $10^{-12}$. We have found that this criterion is sufficient to ensure that the values of the minima to which the relaxation algorithm converges to, reach plateaus to a precision of at least $10^{-10}$.
The whole simulation is performed in double precision and we store the values of the minima in double precision.\footnote{In \cite{Schrock:2012fj} it was shown that the accumulation of numerical errors of typical observables in double precision in lattice gauge fixing is smaller then $10^{-12}$ even for extensively long simulation runs.}
Subsequently we transform each minimum $x\in[0,1.0]$ to an
integer $X\in[0,10^8]$. These integers can then be
unambiguously compared using the bitonic sort algorithm. The inverse of the upper bound of the integer interval defines the resolution with which we choose to distinguish minima.

Hence we consider values of minima as the same when they agree within eight decimal places. 
It is likely that with our resolution of $10^{-8}$ we count some minima as the same which are distinct at finer resolution, i.e.
finer resolution may
eventually allow higher distinction of otherwise non-distinguishable minima. 
Adopting this rather conservative resolution we assure that we obtain real lowest bounds of the number of Gribov copies per orbit which has highest priority for our study.
More details on numerical distinction of Gribov copies, including a discussion of renormalization effects, can be found in \cite{Maas:2011ba}.

In the present work, we sample orbits randomly, i.e., we consider the theory at the strong coupling limit where the inverse coupling
$\beta=0$. This choice of $\beta$ is sufficient to make general conclusions for the number of Gribov copies for the purpose of this work.

\section{Results}

\begin{table*}[htb]
\begin{tabular}{c||c|c|c|c|c||c|c|c||c|c}
 & \multicolumn{5}{|c||}{$d=2$} & \multicolumn{3}{|c||}{$d=3$} & \multicolumn{2}{|c}{$d=4$}\\\hline
$N$  & 2 & 4 & 6 & 8 & 10 & 2 & 4 & 6  & 2 & 4 \\\hline
\# orbits  & 100 & 100 & 100 & 100 & 10 & 100 & 100 & 3  & 100  & 1 \\\hline
\# samples/orbit $[1024^2]$ & 4 & 4 & 4 & 16 & 256 & 4  & 4 & 1024 & 4 & 1024 \\\hline\hline
avg. numb. of minima & 1.1 & 4.7 & 66.2 & 2,464 & 185,709 & 1.8 & 394.7 & $3.335\times 10^6$   & 4.5 & $2.774\times 10^6$ \\\hline
std. dev.            & 0.3 & 2.2 & 35.1 & 1,618 & 110,779  & 0.9 & 228.3 & 79,529.0   & 2.4 & -
\end{tabular}
\caption{The number of orbits and the number of samples per orbit for each lattice size $N^d$ for which we have minimized the gauge functional. The average number of distinct minima that we collected and the corresponding standard deviation is listed.
}
\label{tab:setup}
\end{table*}


\subsection{Number of Gribov copies}\label{sec:nMinimavsSamples}
In Tab.~\ref{tab:setup} we list for each lattice size the number of orbits and the number of samples per orbit for which we have minimized the gauge functional Eq.~(\ref{eq:general_l_g_functional}).
Note that the number of samples is given in units of \emph{mebisample} ($1024^2$ samples).

In Figs.~\ref{fig:nMinima_vs_nSamples_dim2}--\ref{fig:nMinima_vs_nSamples_dim4}
we plot the number of distinct minima that we found as a function of the number of random initial guesses.
Due to the nature of the bitonic sorting algorithm, we measure the number of minima only at stages of powers of two in the number of samples. In the figures, the resulting points are connected by straight lines to guide the eye.

In two dimensions (Fig.~\ref{fig:nMinima_vs_nSamples_dim2}), we find for $N=6$ that all orbits have converged to plateaus, indicating that we are very close to having found all minima in that case. Similarly, the plot for $N=8$ (same figure) reveals that still a relatively large fraction of the orbits have converged. The curves in the plot for $N=10$, in contrast, have not converged and consequently we are further away from having collected all minima here.
Analogously, Fig.~\ref{fig:nMinima_vs_nSamples_dim3} shows the data for $d=3$ where we reach our limit for $N=6$: one gibisample ($1024^3$ samples) per orbit is not sufficient to get close to finding all minima.
In four dimensions, we sample one hundred $2^4$ orbits and a single orbit of $N=4$ with a gibisample initial guesses for which we obtain only a relatively weak lower bound on the number of distinct minima.

We conclude that even though we have not found every single minimum for each orbit, Figs.~\ref{fig:nMinima_vs_nSamples_dim2}--\ref{fig:nMinima_vs_nSamples_dim4} provide clear evidence that the number of Gribov copies is orbit-dependent.

\begin{figure}[htb]
\includegraphics[width=0.5\textwidth]{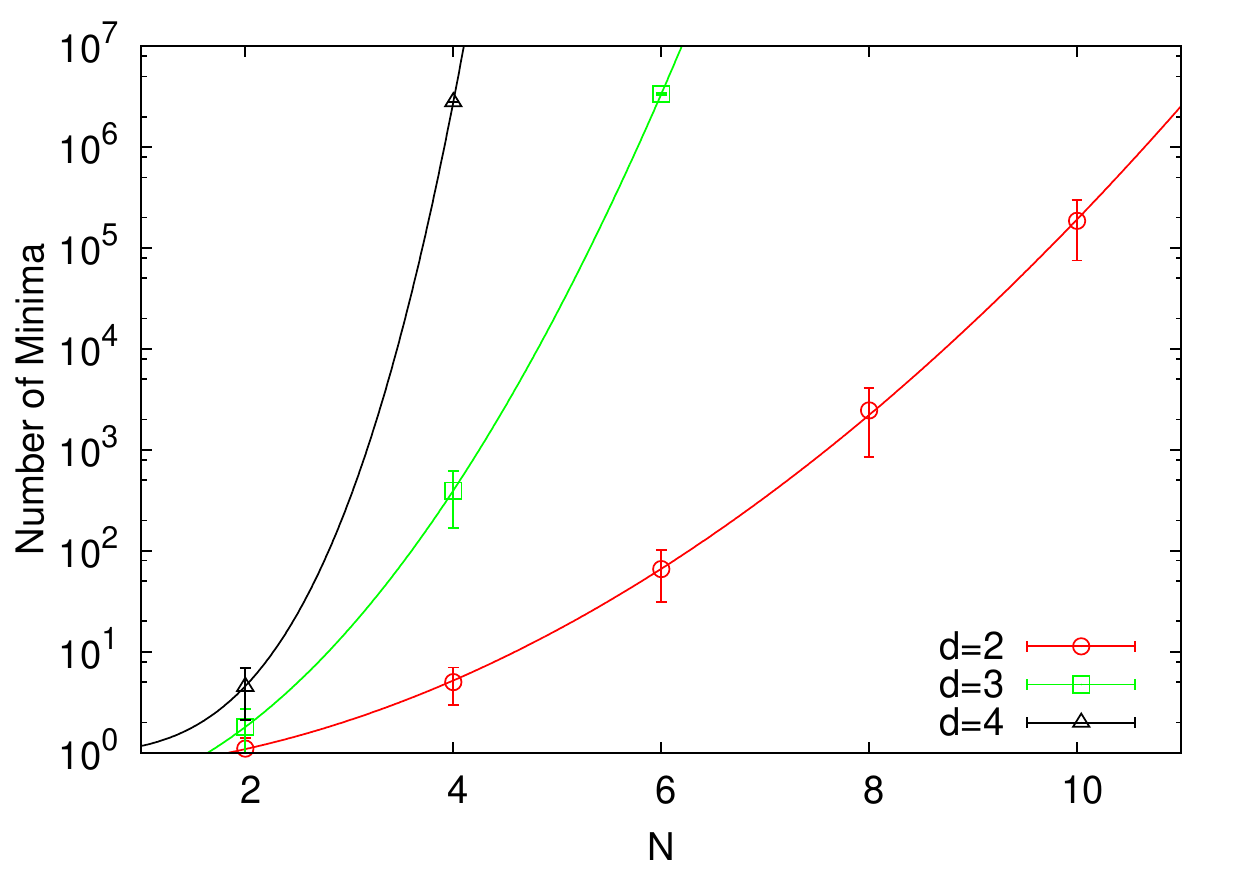}
\caption{Lower bounds of the number of minima as a function of $N$ for $d=2,3,4$ averaged over the different random orbits. 
The curves correspond to best fits to the function Eq.~(\ref{eq:fitfctn}) and the corresponding fit parameters are 
summarized in Tab.~\ref{tab:fitparam}.}
\label{fig:numbMinima}
\end{figure}

\begin{table}[htdp]
\begin{center}
\begin{tabular}{c||c|c|c}
dim & $a$ & $b$ & $c$ \\\hline
2 & 0.631(38) & 0.141(11) & 1.953(33) \\
3 & 0.305(-)  & 0.440(-)  & 2.012(-)  \\
4 & 1 (fixed) & 0.156(-)  & 3.287(-)
\end{tabular}
\end{center}
\caption{The fit parameters of the curves Eq.~(\ref{eq:fitfctn}) shown in Fig.~\ref{fig:numbMinima}.}
\label{tab:fitparam}
\end{table}%

The averages and standard deviations for the lower bounds of the number of minima per orbit and lattice size are summarised in Tab.~\ref{tab:setup}.
The lower bounds on the number of minima as a function of $N$ for all dimensions is plotted in Fig.~\ref{fig:numbMinima}. Additionally, best fits to a function  
\begin{equation}\label{eq:fitfctn}
	h(x)=a\exp\left(bx^c\right)
\end{equation}
are shown and the corresponding fit parameters are listed in Tab.~\ref{tab:fitparam}.
The data for $d=2$ indicate that the number of distinct gauge functional minima depends exponentially on $N^2$. 
The data for $d=3$ do not confirm an exponent $N^d$, but this
is probably because our lowest bound for $6^3$ severely
underestimates the true number of copies.\footnote{The $N=6$ curves in Fig.~\ref{fig:nMinima_vs_nSamples_dim2} are still rising by more then 6\% when increasing the number of samples from 512 to 1024 mebisamples. Moreover, the three curves are rather close to each other compared to, e.g, the curves for $N=4$.}

\begin{table*}[htb]
\begin{tabular}{c||c|c|c|c|c||c|c|c|c||c|c}
 & \multicolumn{5}{|c||}{$d=2$} & \multicolumn{4}{|c||}{$d=3$} & \multicolumn{2}{|c}{$d=4$}\\\hline
$N$  & 2 & 4 & 6 & 8 & 10 & 2 & 4 & 6 & 8 & 2 & 4 \\\hline
\# orbits  & \multicolumn{5}{|c||}{100} & \multicolumn{3}{|c|}{100}  & 20 & 100  & 7\\\hline
avg. glob. min. $F$ 
& 0.431 & 0.249 & 0.219 & 0.210 & 0.208 & 0.441 & 0.338 & 0.328 & 0.324 & 0.482 & 0.413\\\hline
std. dev. 
& 0.139 & 0.037 & 0.022 & 0.016 & 0.014 &0.076 &0.014 &0.006 &0.005 &0.045 &0.006
\end{tabular}
\caption{A complementary set of gauge functional minima: more orbits per lattice size with four mebisamples (4,194,304 samples) per orbit. The averages of the values of the gauge functional at the global minimum and the corresponding standard deviations are listed (cp. Fig.~\ref{fig:glob_min}).
}
\label{tab:setupB}
\end{table*}

\begin{figure}[htb]
\includegraphics[width=0.5\textwidth]{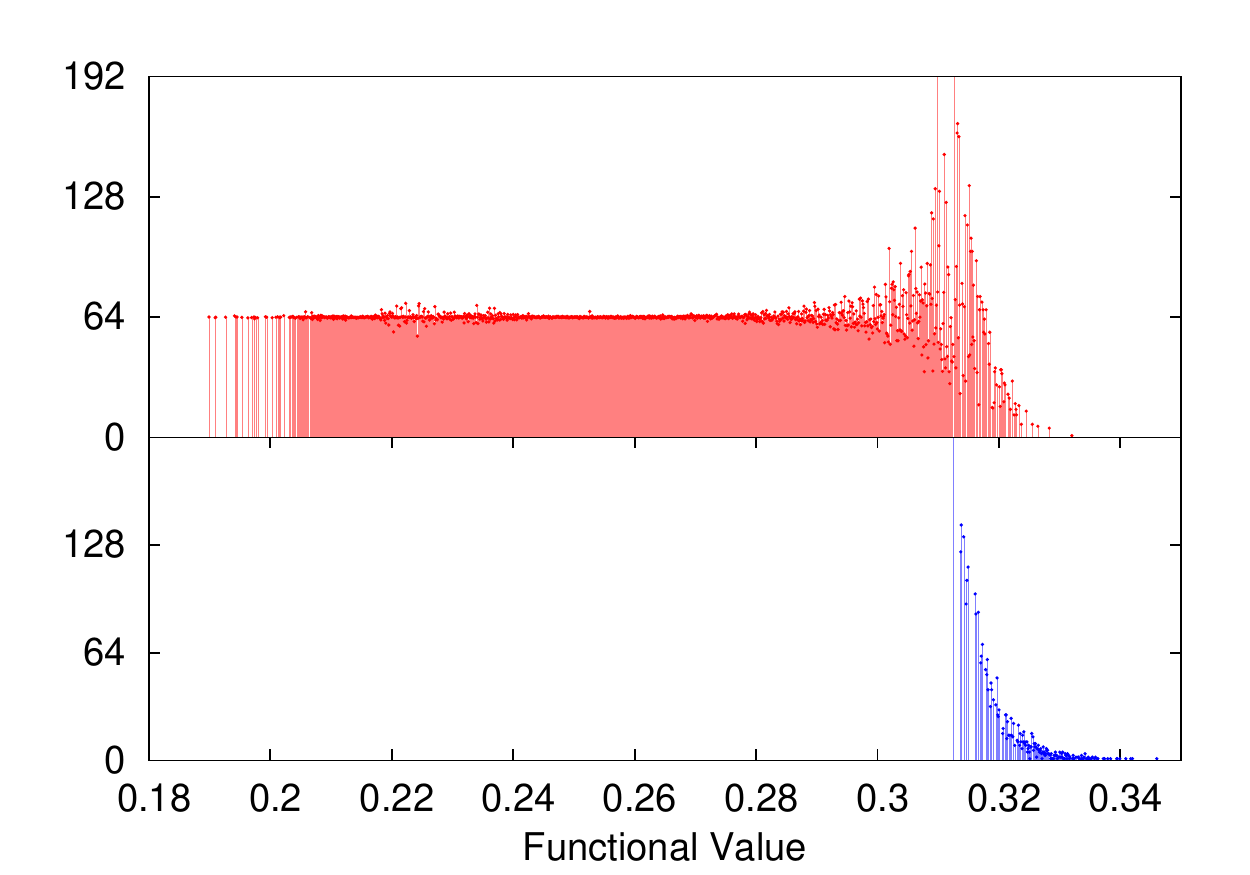}
\caption{In the upper part of the plot we show the ratio of each of the bins (resolution $10^{-4}$) of the distribution of the histograms of the minima of ten orbits of a $10^2$ lattice, taking 256 vs. four mebisamples into account. The lower part of the plot shows the bins from 256 mebisamples when the corresponding bin of four mebisamples was empty (i.e. the ratio in the upper plot was not defined). Points on top of the bin bars are plotted for better visibility.}
\label{fig:histohot_dim2L10_compare}
\end{figure}

\subsection{The values of the gauge functional at the minima and their distribution}\label{sec:attractive}
In order to investigate how many mebisamples we need to sufficiently sample nearly all minima with reasonable statistics, we compare the set of minima we obtained from ten orbits of a $10^2$ lattice from four mebisamples per orbit to the set of minima when we apply 256 mebisample per orbit to the same ten orbits.
We assign each minimum to a bin of resolution $10^{-4}$ and plot the ratios of the entries of the bins from 256 vs. four mebisamples in Fig.~\ref{fig:histohot_dim2L10_compare}. The plot reveals that the ratios of the low-lying and midrange minima is very close to the expected factor $256/4$ whereas the ratios fluctuate much stronger for high-lying minima. Moreover, only very high-lying bins of the four mebisample run are empty while the corresponding bins of the 256 mebisample run have entries, as presented in the lower plot of the figure. This indicates that four mebisamples are sufficient to obtain reasonable statistics and running more samples will improve mainly in the range most distant to the global minimum which appears to be less attractive for the relaxation algorithm.

\begin{figure}[htb]
\includegraphics[width=0.5\textwidth]{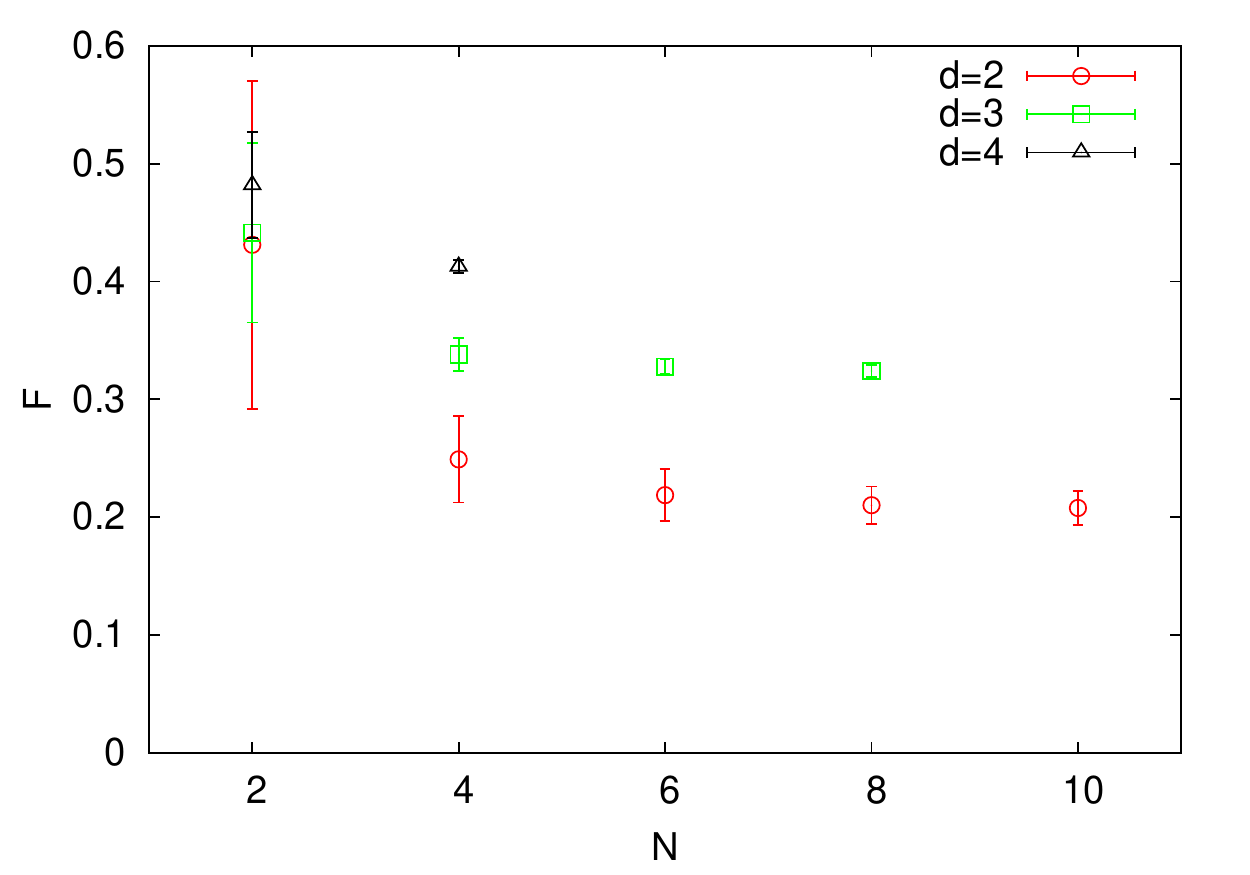}
\caption{The gauge functional evaluated at the global minimum as a function of $N$ for $d=2, 3, 4$.}
\label{fig:glob_min}
\end{figure}

With the aim of studying the dependence of the value of the gauge functional at the global minimum on $N$ and $d$, it is desirable to investigate more orbits to increase the statistics. Motivated by the conclusion of the previous paragraph, we limit the number of samples per orbit to four mebisamples which renders increasing the number of orbits affordable. Nevertheless we are confident that the smallest minimum we find on each orbit is at least numerically very close to the global minimum, if not equal to it. Hence we study the global minima of the orbits listed in Tab.~\ref{tab:setupB} which additionally include $8^3$ lattices.

In Fig.~\ref{fig:glob_min}, the gauge functional Eq.~(\ref{eq:general_l_g_functional})  evaluated at the global minimum as a function of $N$ for $d=2, 3, 4$ is shown.
For constant $N$, the value of the global minimum is higher for higher dimension $d$ and for fixed $d$ the global minima seem to converge to plateaus for $N\gtrsim 6$.

\begin{figure}[htb]
\includegraphics[width=0.5\textwidth]{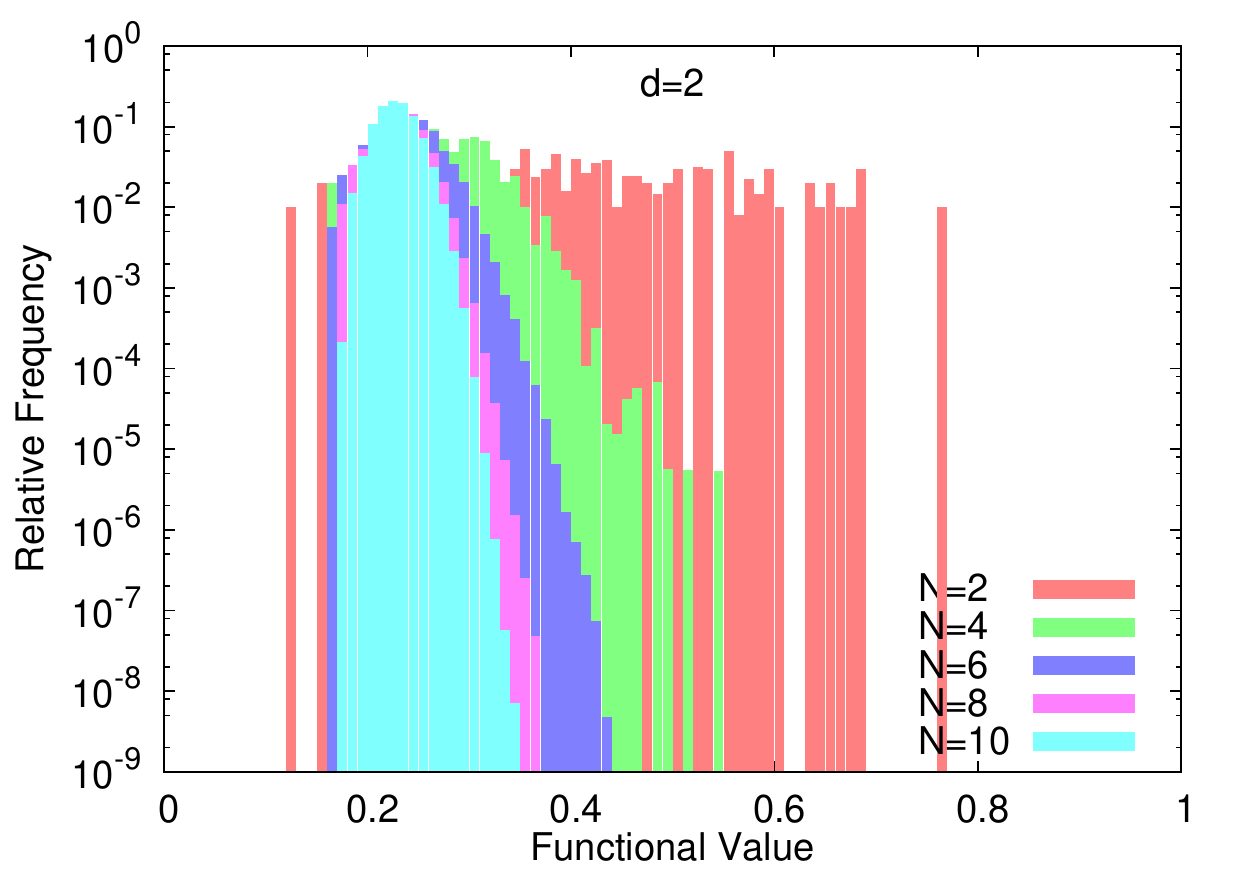}\\
\includegraphics[width=0.5\textwidth]{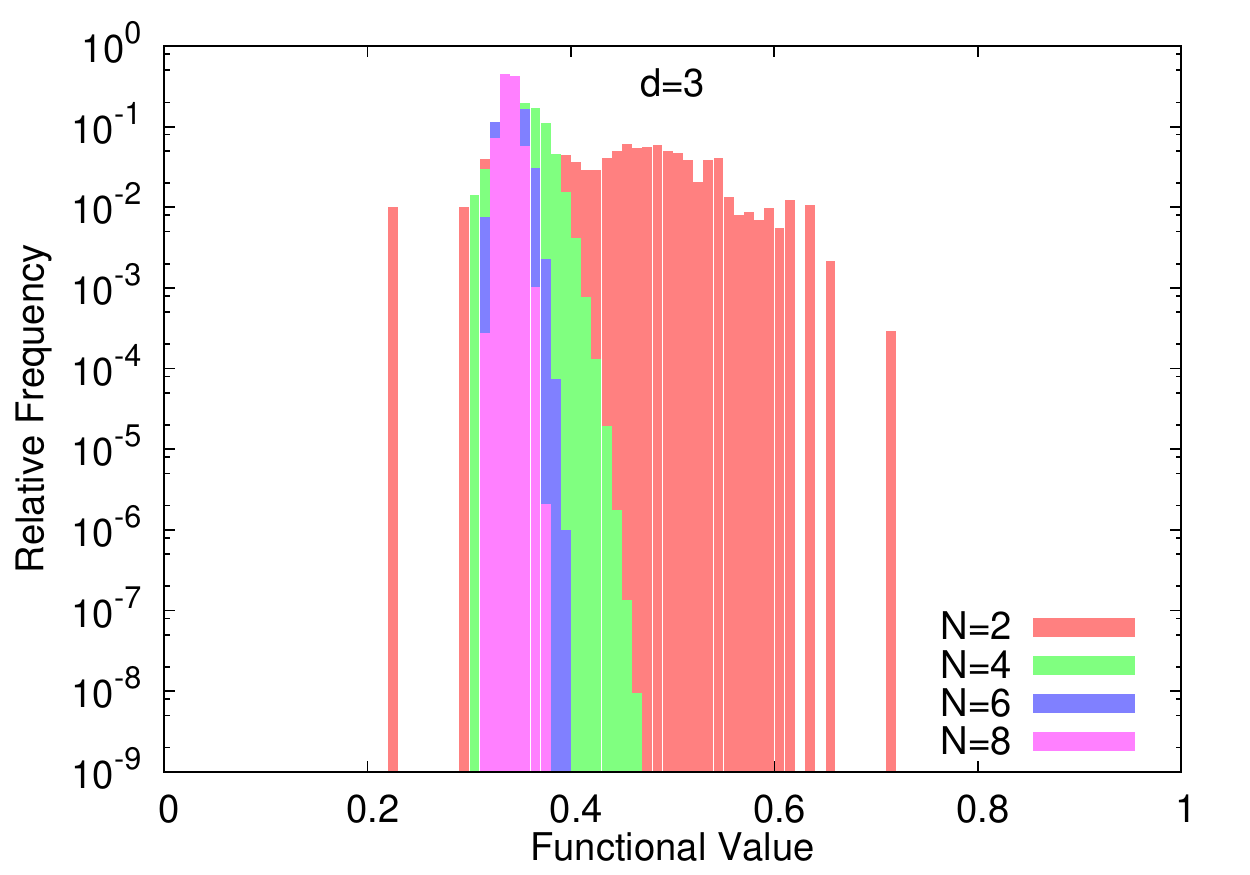}\\
\includegraphics[width=0.5\textwidth]{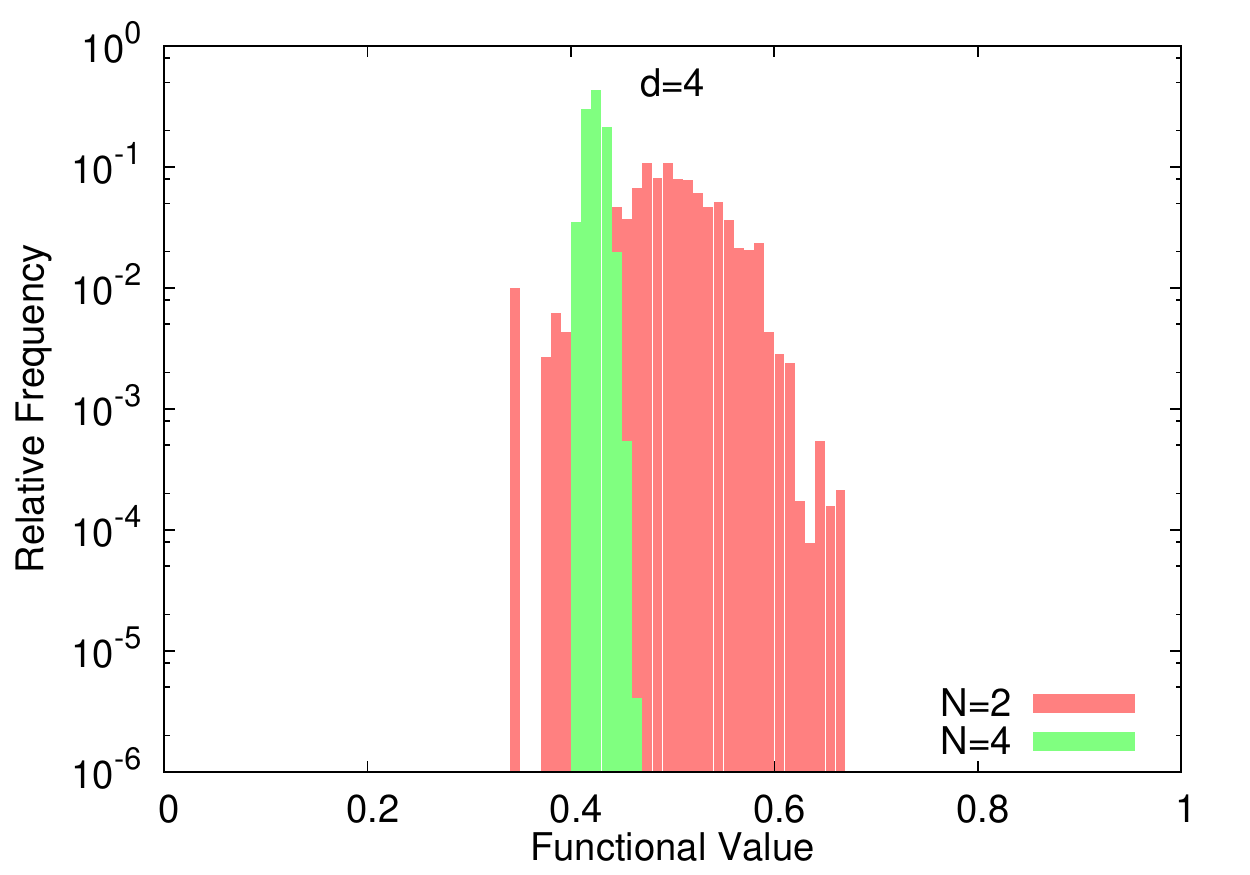}
\caption{The normalised distribution of the minima of all orbits and samples of $d=2, 3, 4$ (see Tab.~\ref{tab:setupB}).}
\label{fig:histohot_dim234}
\end{figure}

\begin{figure}[htb]
\includegraphics[width=0.5\textwidth]{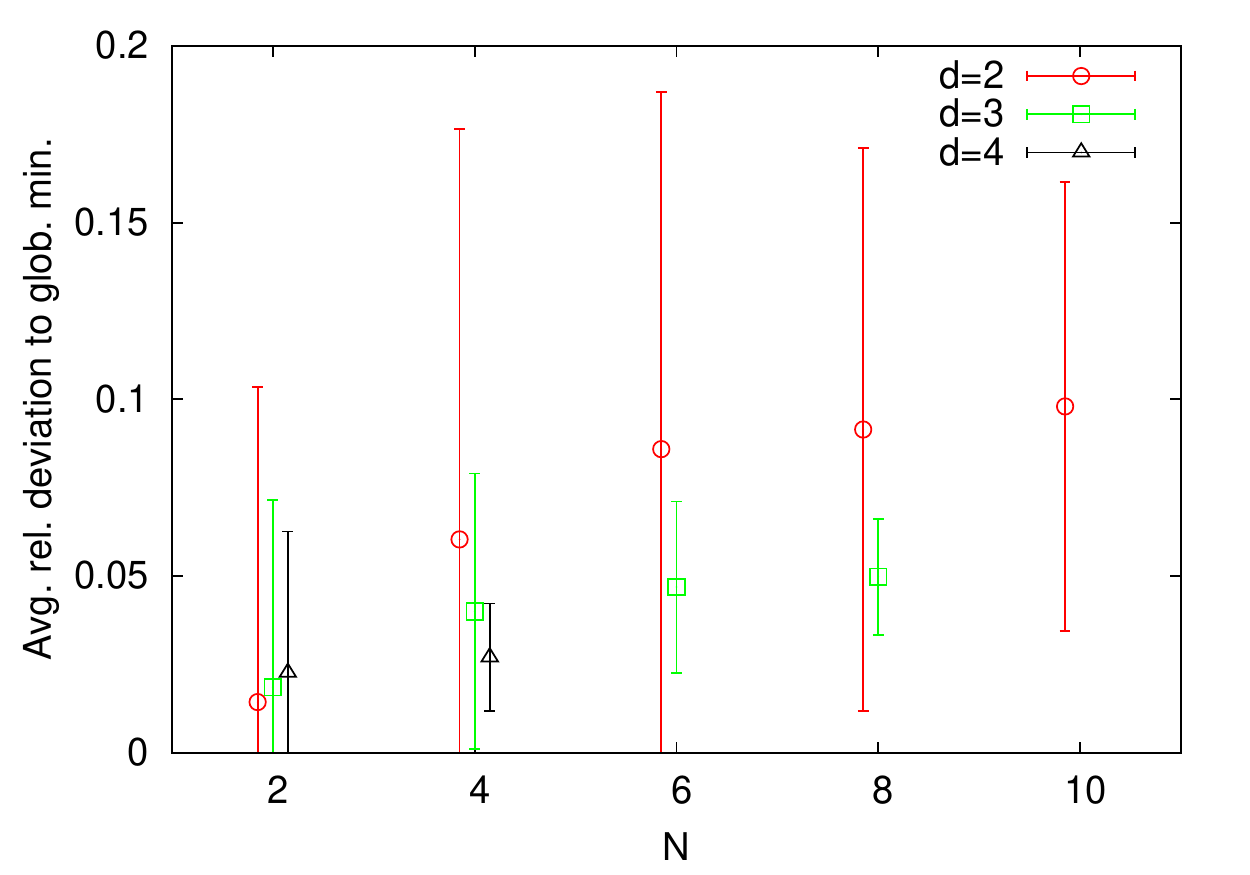}
\caption{The averages and standard deviations of the relative deviation of the local minima $F_i$ from the global minimum $F_g$ on the corresponding orbit: $(F_i-F_g)/F_g$.
}
\label{fig:historelminhot_PBC}
\end{figure}

In Fig.~\ref{fig:histohot_dim234} histograms for
the distribution of the functional values for all lattice sizes, each superimposing data from all orbits of Tab.~\ref{tab:setupB}, are shown.
It is evident that the distribution becomes narrower with increasing lattice size $N$. It is important to stress, however, that the wide spread for lower $N$ is mainly due to the large variance of the value of the global minimum, compare with the data in the table.
Fig.~\ref{fig:historelminhot_PBC}, in contrast, shows the distribution of the minima $F_i$ relative to the global minimum $F_g$ on that orbit: $(F_i-F_g)/F_g$. Subsequently, the data has been averaged over all orbits. As a consequence of this strategy, the aforementioned effect of the variance of the global minima is factored out. The deviation (Fig.~\ref{fig:historelminhot_PBC}) appears to increase with $N$, not decrease as it should if global and local minima became equivalent. In summary, our data does not indicate that global and local minima become equivalent for large $N$ (in the strong coupling limit).


\section{Conclusions}
On the lattice, gauge fixing is formulated as a minimization problem. The stationary points of the gauge fixing functional are the Gribov copies which are physical replications of a gauge configuration and exist even after fixing the gauge non-perturbatively. 
In this paper, we aimed at enumerating the number of Gribov copies in the first Gribov region on the lattice in order to address different modifications of the gauge fixing procedure which can be affected by the potential orbit-dependence of the number of Gribov copies. We studied compact U($1$) gauge theory. The latter not only can serve as laboratory for testing our computational efforts, but is a very important model in its own right: it has been shown that the origin of the Neuberger $0/0$ problem lies in compact U($1$), and the problem is evaded for any SU($N$) when it is evaded for compact U($1$). This holds even though Gribov copies in the compact U($1$) case are just lattice artifacts.

We performed a brute force analysis of the first Gribov region for the compact U($1$) case in $d=2, 3, 4$ dimensions. 
We started the relaxation algorithm from up to more than a billion random points on the gauge orbits and collected up 
to millions of distinct gauge functional minima per orbit.
Even though our GPU implementation has proven to be a powerful tool for counting Gribov copies, we observed that 
the problem of counting Gribov copies becomes increasingly difficult with increasing lattice sizes. In particular, 
for the biggest volumes of our runs ($10^2$, $6^3$ and $4^4$), the convergence to the full number of distinct minima with
an increasing number of minimization attempts (``samples'') could not be achieved. In $d=4$ we reached our limits with a single orbit of the modest lattice size $4^4$.

We were able to show that the number of Gribov copies in the first Gribov region increases exponentially in two, three and four dimensions.
More specifically, we found that the number of distinct minima per orbit increases at least with $\exp\left(\sim N^2\right)$, 
an $\exp\left(\sim N^d\right)$ dependence is likely, though could not definitely be shown with the currently available data.
Moreover, we have found strong indication that the number of minima is orbit dependent, 
i.e., strong indication that the gauge fixing partition function for the minimal Landau gauge on the lattice is orbit dependent.


Finally, in the continuum it was conjectured that the local minima of the corresponding gauge fixing functional tend to be degenerate with the global minimum \cite{Zwanziger:1989mf}. A direct comparison with this conjecture can not be done using our results on the lattice with $\beta=0$. However, while our data exhibits narrowing of the distribution of the values of the gauge fixing functional at the minima when increasing the lattice size (taking data from several orbits per lattice size into account); we cannot observe that local minima tend to get closer to the global minimum of the corresponding orbit.

\begin{acknowledgments}
The authors are very grateful to Reinhard Alkofer, Maartin Golterman, Axel Maas, Jonivar Skullerud, Martin Schaden and Yigal Shamir for helpful discussions and comments on the manuscript.
DM was supported by an ERC and a DARPA Young Faculty Award.
The calculations have been performed on the ``mephisto'' cluster at the University of Graz.
\end{acknowledgments}


%

\end{document}